\documentclass[9pt, journal, compsoc]{IEEEtran}

\usepackage{tikz}

\usepackage{fancyhdr}
\usepackage{graphicx}
\usepackage{subfig}
\usepackage{epstopdf}
\usepackage{pifont}
\usepackage{epsfig}
\usepackage{setspace}
\usepackage{amsmath}
\usepackage{multirow}
\usepackage{floatrow}
\usepackage{array}
\usepackage{comment}
\usepackage{color}
\usepackage{soul}
\usepackage{xspace}
\usepackage{enumitem}
\usepackage{adjustbox}
\usepackage{threeparttable}
\usepackage{textgreek}
\usepackage{dblfloatfix}    

\ifCLASSOPTIONcompsoc
  \usepackage[nocompress]{cite}
\else
  \usepackage{cite}
\fi

\usepackage{array}

\usepackage{setspace}
\ifCLASSINFOpdf
\else
\fi

\begin{document}

\newcommand{\mycomment}[1]{}
\newcommand{\ignore}[1]{}
\newcommand{\code}[1]{\texttt{#1}}
\newcommand{\codesm}[1]{\texttt{\small #1}}
\newcommand{\sm}[1]{{\small #1}}
\newcommand{\mycaption}[2]{\textbf{\textsf{\caption{#1}#2}}}

\newcommand{\hlc}[2][yellow]{ {\sethlcolor{#1} \hl{#2}} }

\newcommand{\todo}[1]{\color{white}\textbf{\hlc[black]{TODO: [#1]}}\color{black}\xspace}
\newcommand{\fixme}[1]{\color{red}\textbf{\hl{FIXME: [#1]}}\color{black}\xspace}
\newcommand{\doit}[1]{\color{white}\textbf{\hlc[red]{DO: [#1]}}\color{black}\xspace}
\newcommand{\remark}[1]{\color{blue}\textbf{\hlc[green]{REMARK: [#1]}}\color{black}\xspace}
\newcommand{\answer}[1]{\color{white}\textbf{\hlc[blue]{ANSWER: [#1]}}\color{black}\xspace}
\newcommand{\pointer}[1]{\color{white}\textbf{\hlc[red]{POINTER: [#1 is working here]}}\color{black}\xspace}
\newcommand{\newedit}[1]{#1\xspace}
\newcommand{\apriori}{\textit{a priori}}
\newcommand{\adhoc}{ad hoc}
\newcommand{\etal}{\textit{et al.}}
\newcommand{\eg}{\textit{e.g.}}
\newcommand{\ie}{\textit{i.e.}}
\newcommand{\mybull}{\noindent $\bullet$~}
\newcommand{\ms}{$ms$~}
\newcommand{\us}{${\mu}s$~}

\newcommand{\mystar}{$\bigstar$}

\newcommand{\figref}[1]{Figure~\ref{#1}}
\newcommand{\sectref}[1]{Section~\ref{#1}}
\newcommand{\sectrefs}[2]{Sections~\ref{#1}-\ref{#2}}
\newcommand{\aref}[1]{Algorithm~\ref{#1}}
\newcommand{\sect}[1]{\section{#1}}
\newcommand{\subsect}[1]{\subsection{#1}}
\newcommand{\subsubsect}[1]{\subsubsection{#1}}
\newcommand{\mysect}[1]{\subsect{#1}}

\newtheorem{ourtask}{Task}

\title{FastDrain: Removing Page Victimization Overheads in NVMe Storage Stack}
\author{Jie Zhang$^{1}$, Miryeong Kwon$^{1}$, Sanghyun Han$^{1}$, Nam Sung Kim$^{2}$, Mahmut Kandemir$^{3}$ and Myoungsoo Jung$^{1}$ \\
{\emph{$^{1}$KAIST, $^{2}$Samsung, $^{3}$PennState}} \\ 
}


\markboth{IEEE COMPUTER ARCHITECTURE LETTER,~Vol.~XX, No.~X, April~2020}%
{Shell \MakeLowercase{\textit{et al.}}: Bare Advanced Demo of IEEEtran.cls for Journals}


\IEEEtitleabstractindextext{%
\begin{abstract}
Host-side page victimizations can easily overflow the SSD internal buffer, which interferes I/O services of diverse user applications thereby degrading user-level experiences. To address this, we propose FastDrain, a co-design of OS kernel and flash firmware to avoid the buffer overflow, caused by page victimizations. Specifically, FastDrain can detect a triggering point where a near-future page victimization introduces an overflow of the SSD internal buffer. Our new flash firmware then speculatively scrubs the buffer space to accommodate the requests caused by the page victimization.   
In parallel, our new OS kernel design controls the traffic of page victimizations by considering the target device buffer status, which can further reduce the risk of buffer overflow.
To secure more buffer spaces, we also design a latency-aware FTL, which dumps the dirty data only to the fast flash pages.
Our evaluation results reveal that FastDrain reduces the $99^{th}$ response time of user applications by 84\%, compared to a conventional system.   
\end{abstract}
}

\maketitle

\IEEEdisplaynontitleabstractindextext
\IEEEpeerreviewmaketitle

\section{Introduction}
\label{sec:intro}
In the past decade, SSDs have successfully replaced spinning disks and become dominant storage media in diverse computing domains, thanks to their performance superiority. 
However, SSDs' access latency is still two orders of magnitude longer than that of the main memory \cite{zhang2019flashgpu}. Since such long latency can significantly degrade the performance of user applications, the host storage stack practically employs a large kernel memory buffer upon the target SSD, called Linux \emph{page cache}.

Even though the page cache to buffer file data can effectively hide the performance penalties in accessing the underlying SSDs, OS kernel often requires to clean the cache by flushing dirty pages to the SSD owing to file synchronization and memory resource depletion. We observe that this cleaning task, called \emph{page victimization}, can severely interfere many legacy I/O requests, issued by other user applications. In practice, modern SSDs in parallel employ a built-in DRAM to buffer multiple incoming write requests, referred to as \emph{internal buffer} \cite{jung2013revisiting}.
However, a large number of dirty page writes on a page victimization can introduce an overflow of the internal buffer thereby significantly degrading user-level experiences. 
Specifically, scrubbing the SSD internal buffer by writing dirty data to the backend flash can block the requests of the user applications from immediate I/O services. This in turn compels the applications to violate a given service level agreement. 

\begin{figure}
\centering
\subfloat[Normalized response time.]{\label{fig:motiv8}\rotatebox{0}{\includegraphics[width=0.45\linewidth]{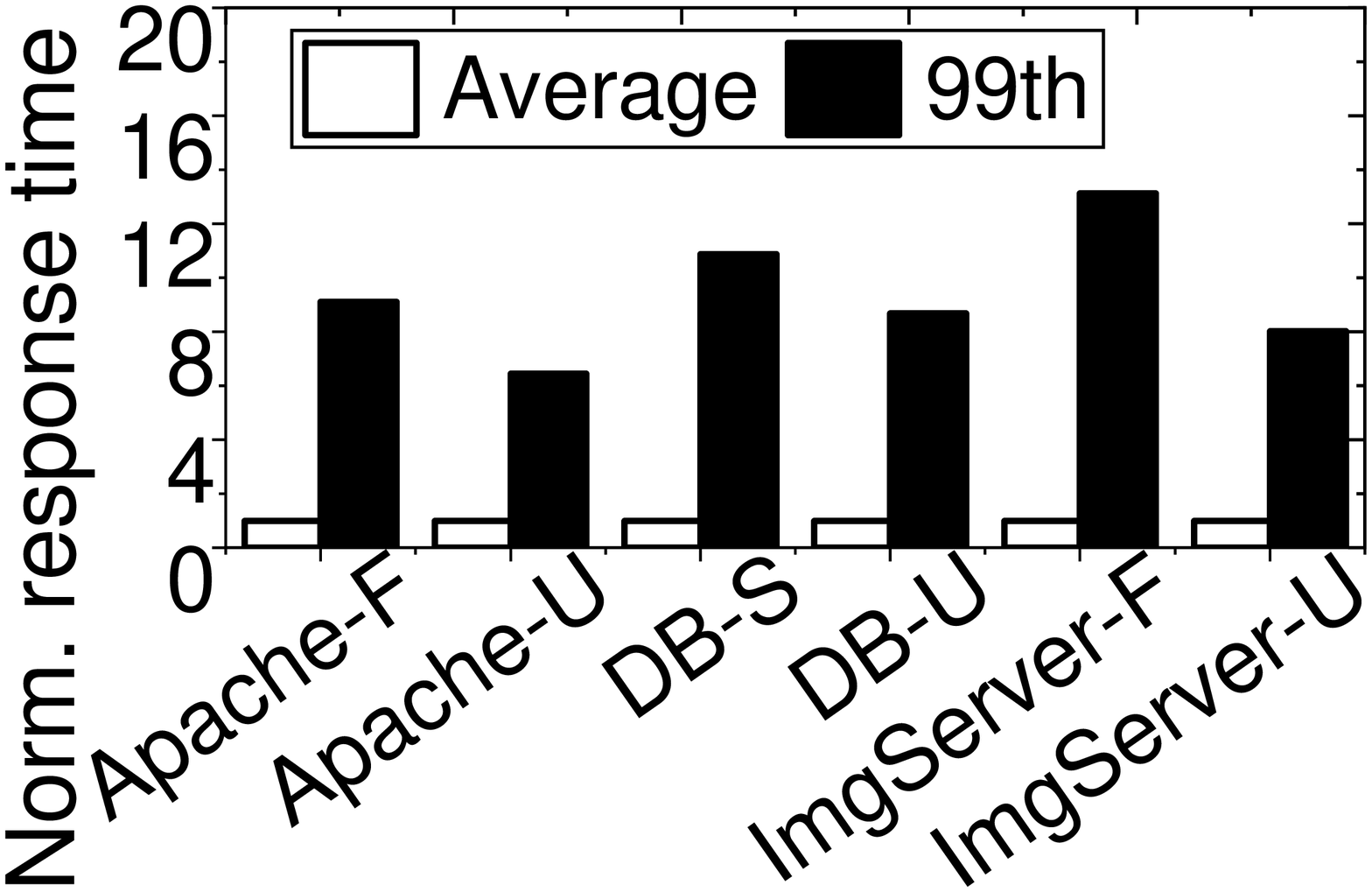}}}
\hspace{1pt}
\subfloat[\textit{Apache-U} response time.]{\label{fig:motiv1}\rotatebox{0}{\includegraphics[width=0.53\linewidth]{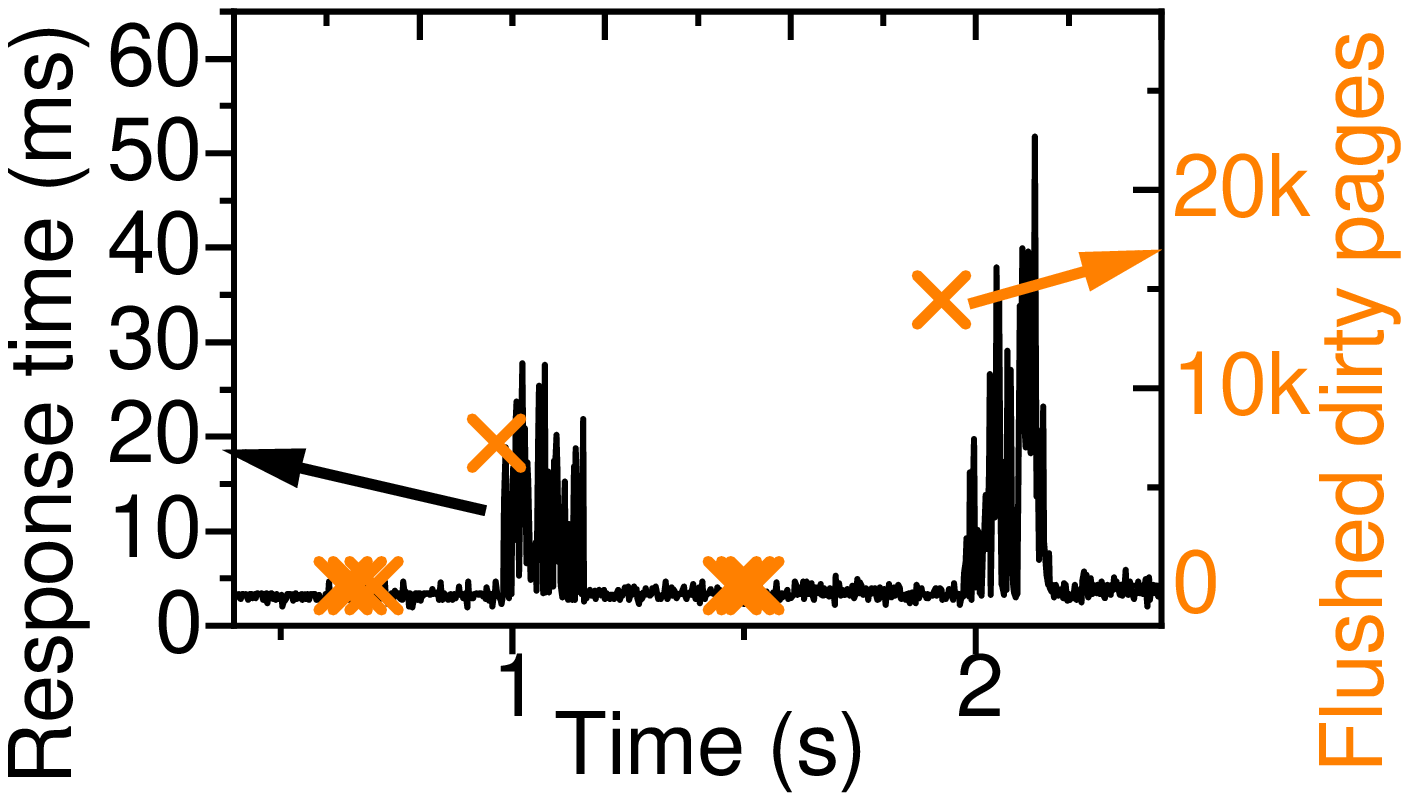}}}
\vspace{-10pt}
\caption{(a) Performance comparison between average and $99^{th}$ response time, and (b) \textit{Apache-U} response time analysis. \vspace{-15pt}}
\label{fig:motiv8_motiv1}
\end{figure}

To quantitatively analyze how the page victimization can affect the user-level experiences, we perform a long-tail latency analysis by running diverse latency-critical applications\cite{APACHE,mysql2001mysql,mosberger1998httperf} together with a throughput-oriented application\cite{gailly2010gnu} (cf. Section \ref{sec:evaluation} for experiment details). Figure \ref{fig:motiv8} shows the results that compare the average latency with $99^{th}$ \emph{response time} of the corresponding latency-critical applications. The $99^{th}$ response time of all latency-critical applications is 9 $\times$ longer than their average response time, in overall.
To understand the root cause of the long $99^{th}$ response time, we also study the execution behavior of a representative application, \textit{Apache-U}. The results are shown in Figure \ref{fig:motiv1}. In addition to a time series analysis of \textit{Apache-U}'s response time, this figure includes the number of dirty pages, flushed by the page cache. \textit{Apache-U}'s response time drastically increases when the OS kernel starts the process of heavy page victimizations.
This is because the OS kernel flushes the dirty pages in the page cache without understanding the status of the underlying SSD, which results in an overflow of the internal buffer. 
Specifically, this buffer overflow enforces flash firmware to write a bulk of dirty data from the built-in DRAM to the flash, which makes the SSD backend too busy to serve other legacy I/O requests (coming from \textit{Apache-U}). The I/O requests are suspended until the SSD backend is available. Consequently, the following data processing is postponed due to this I/O service suspension, which in turn degrades the user experience of \textit{Apache-U}.

In this work, we propose \emph{FastDrain} that coordinates both OS kernel and flash firmware to address the long-tail latency issue, caused by the page victimizations. 
Specifically, the OS kernel signals the flash firmware about an incoming page victimization event via a PCIe message. Our new flash firmware design leverages the message to speculatively allocate free spaces in the built-in DRAM, which can immediately absorb the incoming heavy loads (imposed by the page victimization). In parallel, the flash firmware notifies the OS kernel when the page victimization causes the buffer overflow in the near-future. To this end, we modify the OS kernel to adaptively control the traffic of page victimizations by using an upcall message. Note that the internal buffer has a relatively small DRAM size, which cannot accommodate all the requests, introduced by the host-side page victimization. Thus, we also design a latency-aware FTL (\emph{flash translation layer}) at the device level. This FTL enables the internal buffer to secure as many DRAM spaces as possible by quickly dumping out the dirty data to only \newedit{fast LSB flash pages}, which can reap the benefits of reducing the interference impact for applications' requests.
Our evaluation results show that FastDrain reduces the $99^{th}$ response time by 84\% in comparison to a conventional system. 

\section{Background}
\label{sec:background}

\begin{figure}
\centering
\def\subfigcapskip{0pt}
\subfloat[I/O storage stack.]{\label{fig:back_design}\rotatebox{0}{\includegraphics[width=0.47\linewidth]{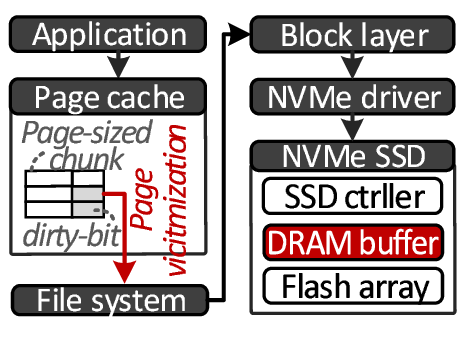}}}
\hspace{1pt}
\subfloat[\newedit{Performance comparison.}]{\label{fig:motiv5}\rotatebox{0}{\includegraphics[width=0.51\linewidth]{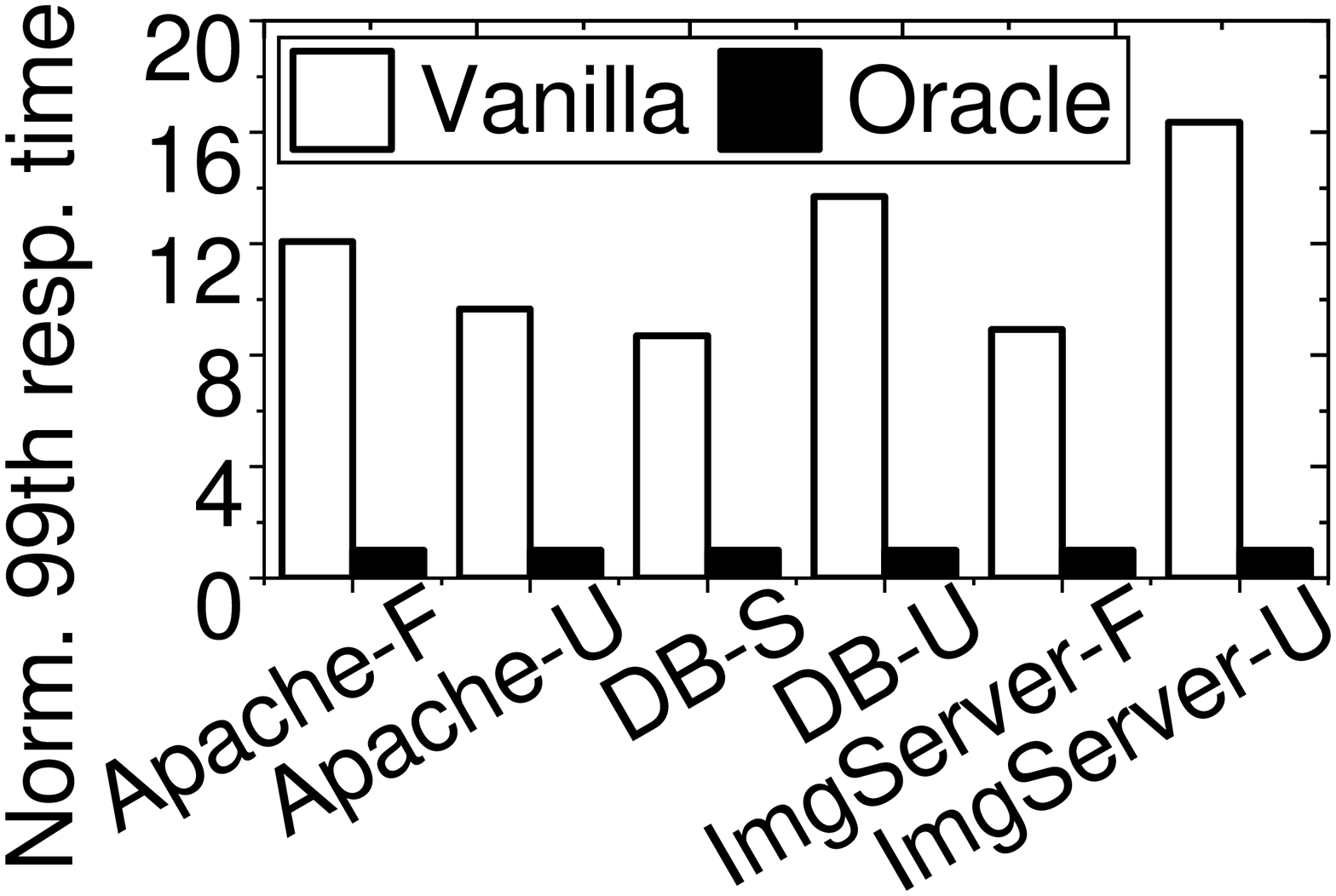}}}
\vspace{-10pt}
\caption{\newedit{(a) I/O storage stack, and (b) $99^{th}$ response time of \texttt{Vanilla} and \texttt{Oracle} systems (normalized to \texttt{Oracle})}. \vspace{-15pt}}
\label{fig:back_motiv5}
\end{figure}

Figure \ref{fig:back_design} illustrates the storage stack existing from a user process to low-level flash media. When the user process issues I/O requests, the requests are forwarded to Linux page cache and file system modules. \newedit{If the target data does not exist in the page cache, an I/O scheduler (e.g., kyber \cite{kyber}) in the Linux multi-queue block layer reorders the read and write requests by prioritizing the read requests. The requests are then passed through the NVMe driver and finally served by the SSD.}

\noindent \textbf{Page cache.} The host employs the page cache to cache frequent-accessed data chunks of files, which can speed up the file accesses. 
However, the OS kernel needs to frequently flush dirty pages from the page cache to the SSD to satisfy the data durability guarantee and/or secure enough DRAM space for the incoming I/O requests. 
The flush activities of page victimization can be in practice categorized as \emph{background} and \emph{foreground} tasks. Specifically, the flush task is executed along with the user applications as background activities when the number of dirty pages is greater than a low threshold (denoted by the \texttt{dirty\_background\_ratio}) or when the dirty pages have a period of life longer than a timer that the user configures. Otherwise, the OS can suspend user processes and prioritize the page victimization as foreground task when the system call of \texttt{fsync()} is invoked or when the number of dirty pages is greater than a high threshold (denoted by the \texttt{dirty\_ratio}).   

\noindent \textbf{SSD internal.}
Modern SSDs typically consist of a SSD controller (flash firmware), built-in DRAM modules and a large number of flash packages. To increase the storage throughput, the SSDs employ multiple \emph{channels}, each connecting to a number of flash packages over a flash system bus (i.e., ONFI \cite{workgroup2011open}). Each flash package also contains multiple flash \emph{dies}. Flash dies across all flash packages can simultaneously serve different memory requests, which exposes a high degree of SSD internal parallelism.
\newedit{However, writing data to flash takes access times much longer than those of read operations. Typically, the write latency varies ranging from 560 \textit{us} to 5 \textit{ms} based on which type of flash pages being accessed. Table \ref{tab:gem5} shows the write variation observed by different types (LSB/CSB/MSB) of flash pages, which are extracted from a real 25 \textit{nm} flash sample \cite{zhang2015opennvm}. Due to the long write latencies, writing flushed dirty pages of page victimization to flash can, unfortunately, make the target SSD backend too busy to serve other legacy requests being issued by user applications. To address this, most SSDs employ the built-in DRAM modules as an internal buffer and accommodate the incoming write requests before directly writing them to the backend flash media. Since the SSD built-in DRAM buffers the flushed dirty pages of page victimization, the SSD backend is available to serve the applications' requests.} However, the internal buffer usually has a limited space, which is unable to buffer a bulk of dirty pages from a heavy page victimization.

\section{Eliminate Page Victimization Overhead}
\label{sec:highlevelview}

\subsection{Overview}
\noindent \textbf{Co-design of page cache and SSD internal buffer.}
To prevent the page victimization from interfering the I/O services of the user applications, one solution is to coordinate both OS kernel and flash firmware to avoid the overflow of the small SSD internal buffer. 
However, it is challenging to enable such coordination due to the lack of communication between them: the existing flash firmware is unaware of the host-side page victimization, while the OS kernel is completely disconnected from the device-level buffer status.
To address this, we enable a message passing across the storage stack. Specifically, we revise Linux I/O service routine that delivers I/O requests from a user to a physical device by piggybacking useful information onto the requests. We also modify the NVMe driver and controller, which can enable all software/hardware modules in the OS and SSD to communicate in a full-duplex manner.
When an incoming page victimization event is detected from the page cache, a notification is passed through to the flash firmware. This information allows the flash firmware to optimize its internal buffer to accommodate the flushed dirty pages of page victimization, which can mitigate the interference to the I/O accesses of the user applications. In addition, the flash firmware reports the status of its internal buffer to the OS kernel when the buffer is nearly full. Using this upcall report, the OS kernel can precisely control the traffic of page victimization based on the device-level buffer status.  

\noindent \textbf{Page victimization aware FTL.}  
While the co-design of OS kernel and flash firmware can improve the efficiency of the internal buffer, some write requests should be served by the flash media as soon as possible to alleviate I/O interference and congestion. For example, data eviction of the SSD internal buffer needs to be quickly drained if the buffer is almost full and an activity of the page victimization (coming from the page cache) is detected. \newedit{To appropriately handle such cases, we design a new FTL, which serves the urgent requests of the foreground data eviction in fast LSB pages residing across the flash blocks of all existing flash dies. This design not only shortens the write latency but also improves its internal parallelism. We also accommodate background (non-urgent) data eviction in slow CSB/MSB flash pages to balance the writes in different types of flash pages, thereby utilizing as many flash pages as possible in the SSD.}

\begin{figure}
\centering
\def\subfigcapskip{0pt}
\subfloat[\newedit{Impact of flash write latencies.}]{\label{fig:motiv6}\rotatebox{0}{\includegraphics[width=0.48\linewidth]{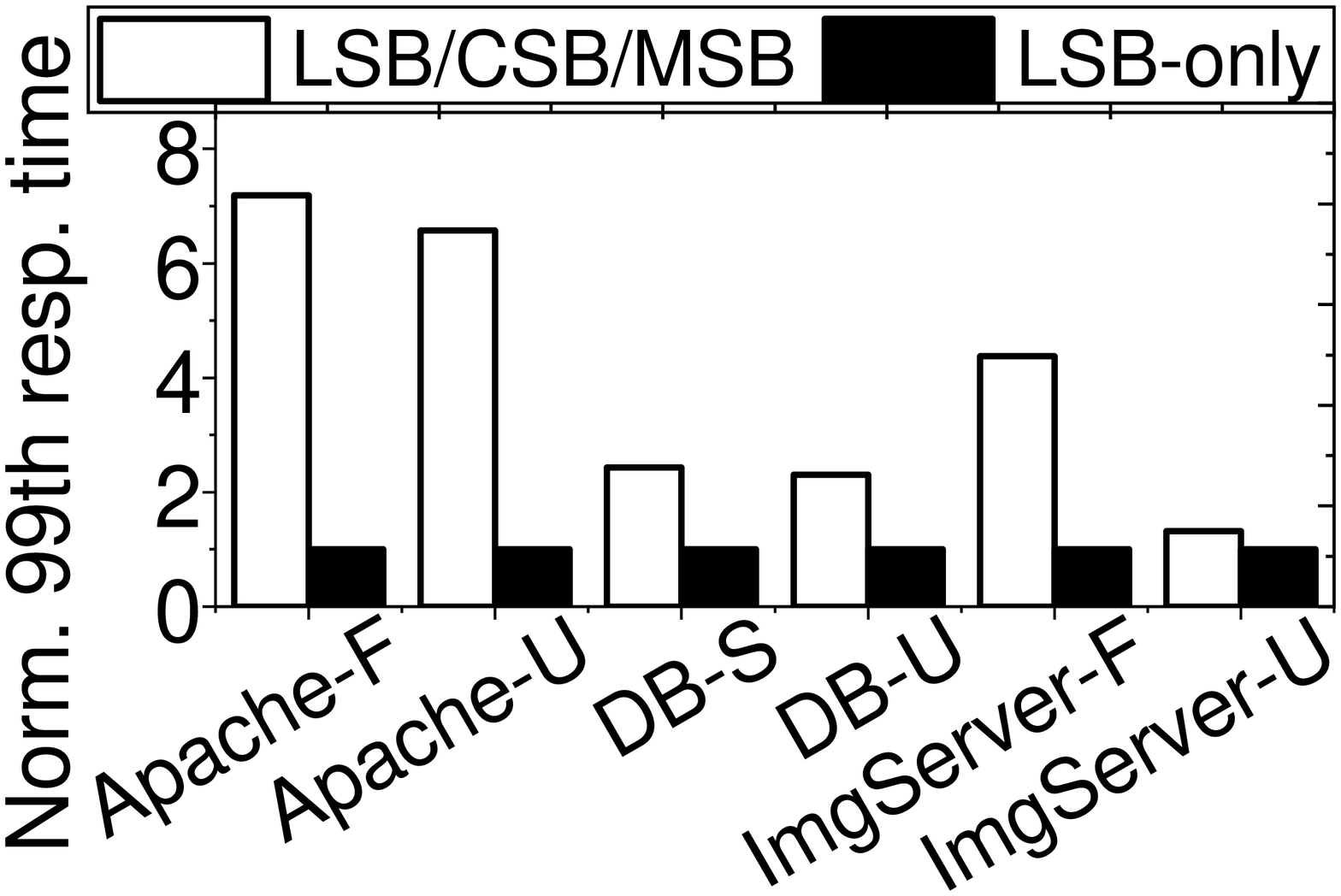}}}
\hspace{3pt}
\subfloat[Flash page wastage.]{\label{fig:motiv7}\rotatebox{0}{\includegraphics[width=0.48\linewidth]{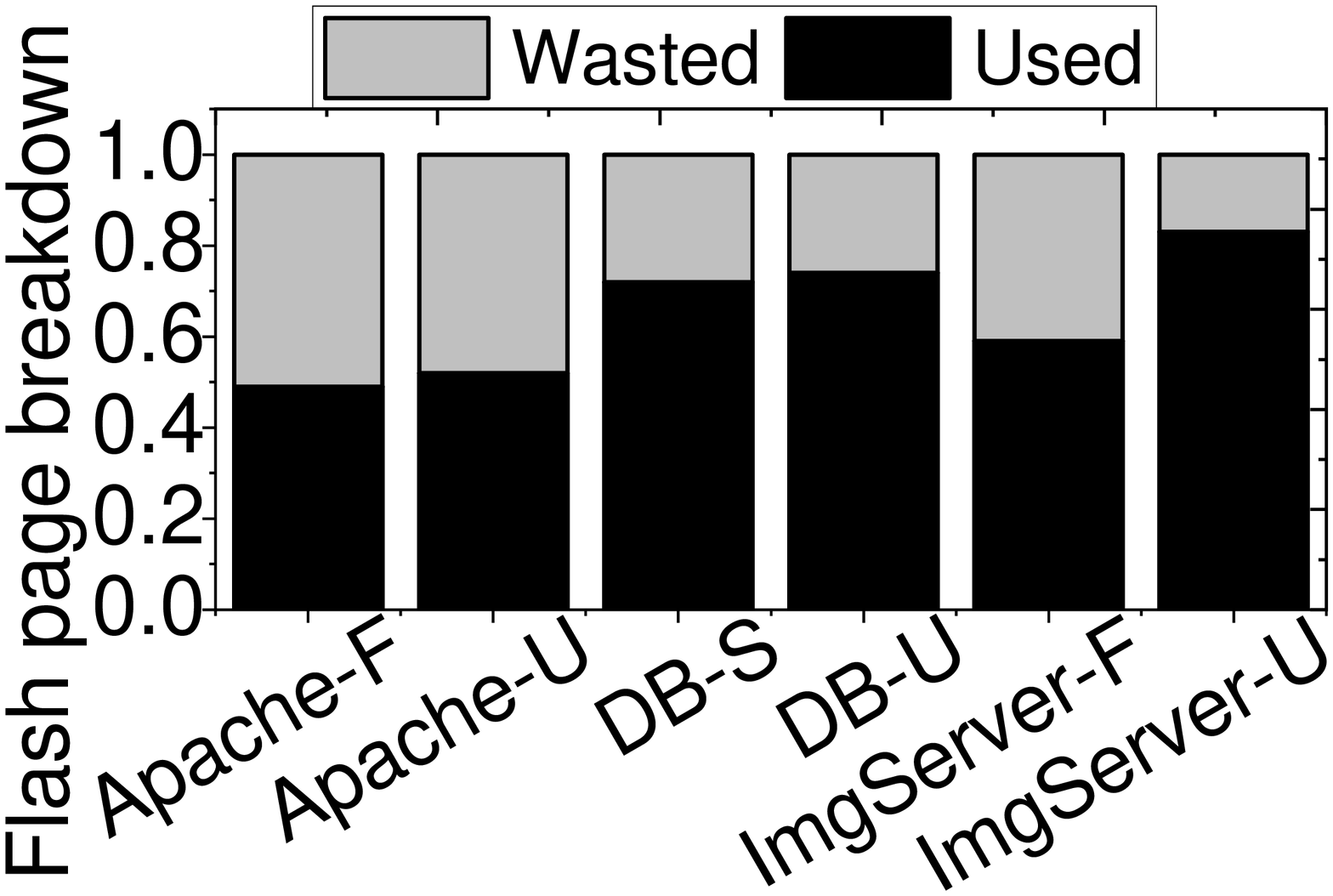}}}
\vspace{-10pt}
\caption{\newedit{(a) The impact of flash write latencies on $99^{th}$ percentile response time (normalized to \texttt{LSB-only}), and (b) flash page usage analysis of the prior work.} \vspace{-15pt}}
\label{fig:back_motiv5}
\end{figure}


%


\subsection{Holistic Optimization of NVMe Stack}
\label{sec:storage}
\noindent \textbf{SSD internal buffer management for page victimization.}
We measure the impact of the SSD internal buffer on user-level experiences by comparing the performance of two SSDs: a traditional SSD (\texttt{Vanilla}) and an oracle SSD with an infinite SSD internal buffer (\texttt{Oracle}). The evaluation results on various applications are shown in Figure \ref{fig:motiv5}. By completely buffering dirty data in its buffer, \texttt{Oracle} can prevent the page victimization from blocking applications' I/O services, thereby reducing the $99^{th}$ response time of the applications by 91\%, on average.

Motivated by this, we design foreground and background buffer eviction strategies in flash firmware to improve the efficiency of the internal buffer: the foreground eviction strategy can clean up the buffer space to accommodate the page victimization as much as possible, while the background eviction strategy smoothly evicts buffered dirty pages to the underlying storage with a minimal penalty of blocking applications' I/O services. We introduce \textit{high\_threshold} and \textit{low\_threshold} as the watermark of the buffer usage and adjust different eviction strategies based on the runtime buffer usage.
Specifically, our new flash firmware can estimate the future buffer usage by accounting the number of dirty pages in the incoming page victimization.
If the internal buffer has a sufficient space (the future buffer usage is lower than a high\_threshold), the background eviction strategy is triggered. 
The flash firmware then estimates the average number of flash channels/dies that are intensively accessed by the applications' I/O requests. For each internal buffer eviction, we evict a limited set of pages to the flash dies in idle.
On the other hand, if the internal buffer is insufficient to serve incoming I/O requests related with the page victimization (the future buffer usage is higher than a high\_threshold), our foreground eviction strategy actively evicts dirty pages to all flash dies until the SSD internal buffer usage is less than low\_threshold.


\noindent \textbf{Latency-aware FTL design.}
The foreground eviction can severely block applications' I/O requests, as it actively dumps a large number of dirty pages to flash. To mitigate the disturbance from the foreground eviction, one potential solution is to selectively write/program the foreground eviction in fast LSB pages and write the background eviction sequentially in LSB/CSB/MSB pages. 
Figure \ref{fig:motiv6} shows the performance benefits, brought by our solution, in executing latency-critical applications. 
As shown in the figure, programming foreground eviction in fast flash pages only (\texttt{LSB-only}), on average, can reduce the $99^{th}$ response time by 65\% under all workloads, compared to programming foreground eviction sequentially in all types of flash pages (\texttt{LSB/CSB/MSB}).    
To make FTL be aware of the fast pages, prior work \cite{grupp2009characterizing} maintains a ``write point'' to find out the target LSB flash page. For example, if the write point marks the next available page as a MSB page in the current flash block, the FTL can skip the MSB page and place dirty data in the next LSB page. 
However, as foreground eviction simultaneously writes a large number of dirty pages to flash, prior work has to skip lots of CSB/MSB pages, which unfortunately wastes flash spaces. Our evaluation results show that 35\% of flash pages are underutilized in the evaluated workloads, due to the foreground eviction (cf. Figure \ref{fig:motiv7}).

To reduce the write latency and improve flash utilization, we develop a new \emph{latency-aware} FTL design. Our design groups the LSB pages across all available flash blocks to accommodate the dirty data of foreground eviction without skipping CSB/MSB pages. In addition, we interleave the foreground and background evictions to access LSB and CSB/MSB pages, respectively, to balance the flash page usage.
To manage the available resources in the flash blocks, we allocate a \emph{block pointer} for each flash block, which points to the free page for program operations. All block pointers are categorized into three groups based on the types of free pages: LSB, CSB and MSB page groups. Within each page group, the block pointers belonging to the same flash die are organized as a list and all lists are indexed by the flash die ID. 
To accommodate dirty pages from a foreground eviction, our proposed FTL selects the flash blocks from different lists in the LSB group for free LSB pages. Therefore, the dirty data are interleaved to the LSB pages across different flash dies for better parallelism. Similarly, our FTL prefers choosing the flash blocks in CSB and MSB page groups to store the dirty data from background eviction. If CSB and MSB pages are unavailable, flash blocks in LSB page groups will be selected for background eviction. Note that the unused flash space (i.e., CSB/MSB pages) due to foreground eviction can be reclaimed by garbage collection, which would not hurt the total SSD capacity. We also constrain the maximum LSB-only flash space (cf. Table \ref{tab:gem5}) to keep high effective SSD capacity.

\begin{figure}
	\centering
	\vspace{-5pt}	
	\includegraphics[width=1\linewidth]{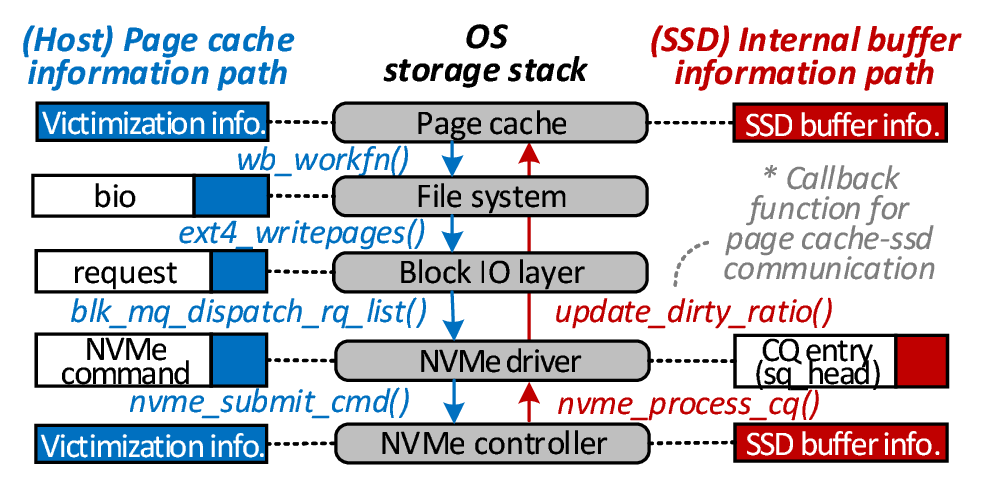}
	\vspace{-30pt}
	\caption{Page cache and SSD communication. \vspace{-20pt}}
	\label{fig:impl_stack}
\end{figure}

\noindent \textbf{Adaptive page victimization scheduling.}
It is undesirable to face a situation that the SSD internal buffer is close to full, while the OS kernel is still flushing dirty pages in the page cache. 
Although our FTL design can accelerate the draining procedure of the SSD internal buffer, intensive foreground eviction can overuse LSB flash pages, which wastes the storage space. To address this, we propose to adjust the traffic of page victimization in OS kernel based on the status of the SSD internal buffer. Specifically,
we configure two sets of \texttt{dirty\_ratio} and \texttt{dirty\_background\_ratio} in the page cache, which have high (10\% and 5\% in this paper) and low values (5\% and 3\%). By default, the set of low dirty ratio values is applied in the page cache to reduce the occupancy of memory spaces. If the flash firmware reports that its internal buffer is occupied higher than the high threshold, the OS kernel stops flushing the dirty pages by temporarily using the set of high dirty ratio values. Once the SSD reports that its internal buffer utilization has fallen back to the low threshold, the OS kernel adopts the set of low dirty ratio and flushes the dirty pages to the SSD.

\subsection{Page Victimization Awareness}
\noindent \textbf{Page victimization detection.}
It is non-trivial for the flash firmware to speculate the page victimization events sololy based on the patterns of incoming I/O requests. Since OS kernel manages the page cache, we propose to extract the information of page victimization by monitoring every flush event at runtime and checking the number of dirty pages to be flushed.
Specifically, to manage the page victimization, OS kernel maintains a pool of working threads, called \texttt{bdi\_writeback\_threads}, each of which can be invoked to process the flush tasks. By monitoring the actions of \texttt{bdi\_writeback\_threads}, we can detect each event of page victimization. For every page victimization, we check the number of pages to be flushed by leveraging the data structure \texttt{bdi\_writeback}. \texttt{bdi\_writeback} is created to maintain a list of dirty inodes for flushing. By walking through the list of dirty inodes, we can obtain the number of dirty pages.

\noindent \textbf{Bidirectional datapath for communication.}
To leverage the extracted information of page victimization and device buffer status, one potential issue is the lack of data path between the page cache and the SSDs for communication.
We set up the datapath by adopting the tagging methods in storage system \cite{zhang2018flashshare,mesnier2011differentiated}. Specifically, we tag the page victimization information and SSD internal buffer status in an NVMe command and an NVMe completion message, respectively. 
Figure \ref{fig:impl_stack} shows the details of our proposed datapath. Specifically, page victimization needs to pass through the file system, block I/O layer, NVMe driver, and NVMe controller. Once the storage-side NVMe controller receives the NVMe commands, it extracts the page victimization information and informs the flash firmware. Similarly, the SSD internal buffer status can be embedded in the NVMe completion message by overriding a field called \texttt{sq\_head}.
Figure \ref{fig:impl_stack} shows the datapath for transferring the SSD internal buffer information (i.e., full) from the SSD to the host. When the NVMe driver extracts SSD internal buffer information in the function, \texttt{nvme\_process\_cq}, the OS kernel modifies global variables of flush thresholds \texttt{dirty\_ratio} and \texttt{dirty\_background\_ratio}.

\section{Evaluation}
\label{sec:evaluation}



\subsection{Experimental Setup}
\noindent \textbf{Experiment environment configuration.}
We leverage Amber \cite{gouk2018amber}, a full-system simulator, to perform precise simulation on the SSD hardware and the existing software from OS to SSD firmware. 
In our experiment, the storage is an NVMe SSD instance with TLC flash \cite{zhang2015opennvm}.  
We also employ 512MB DDR3 DRAM as SSD internal buffer, which is similar to Intel 750 NVMe SSD \cite{NVMeSSD}. 
Our simulation details are shown in Table \ref{tab:gem5}.

\begin{table}
\centering
\resizebox{\columnwidth}{!}{
\begin{tabular}{|l|c|l|c|}
\hline
\multicolumn{1}{|c|}{\textbf{Host}} & \textbf{Values} & \multicolumn{1}{c|}{\textbf{Storage (SSD)}} & \textbf{Values} \\ \hline
\multirow{3}{*}{\textbf{Processor}} & ARM v8 & \textbf{Read(LSB/CSB/MSB)} & 58/78/107us \\ \cline{2-4} 
 & 8 core & \textbf{Write(LSB/CSB/MSB)} & 0.56/2.2/5ms \\ \cline{2-4} 
 & 2GHz & \textbf{Erase/Channel/Package} & 2.27ms/16/4 \\ \hline
\textbf{L1I/L1D} & 64KB/64KB & \textbf{Die/Plane/Page size} & 2/2/8KB \\ \hline
\textbf{L2 cache} & 2MB & \textbf{Capacity/Internal buffer} & 800GB/512MB \\ \hline
\textbf{Mem ctrler} & 1 & \textbf{High/Low thresholds} & 0.8/0.2 \\ \hline
\textbf{Memory} & DDR3, 4GB & \textbf{Max LSB-only region} & 64GB \\ \hline
\end{tabular}}
\vspace{-15pt}
\caption{System configuration parameters. \label{tab:gem5} \vspace{-20pt}}
\end{table}


\noindent \textbf{System configurations.}
We implement three different computer systems by adding the proposed optimization techniques and compare them against a vanilla system.

\begin{enumerate}[leftmargin=8pt, itemsep=-1ex,topsep=0.7ex,partopsep=0.7ex,parsep=1ex]

\item \textbf{Vanilla}: A Vanilla system running with NVMe SSD;

\item \textbf{FD-Buf}: \newedit{Integrating our SSD internal buffer management design into \texttt{Vanilla}, which can adaptively perform SSD buffer eviction based on page victimization events;}

\item \textbf{FD-FTL}: \newedit{Introducing our latency-aware FTL into \texttt{FD-Buf}, which selectively programs foreground eviction to LSB pages;}  

\item \textbf{FD}: \newedit{Integrating our adaptive page victimization scheduling scheme into \texttt{FD-FTL}, such that Linux page cache can suspend evicting dirty pages when the SSD internal buffer is full.} 

\end{enumerate}

\noindent \textbf{Workloads.} 
\newedit{We evaluate six representative latency-critical workloads: \textit{Apache-F}, \textit{Apache-U}, \textit{DB-S}, \textit{DB-U}, \textit{ImgServer-F} and \textit{ImgServer-U}, which are obtained from \cite{APACHE,mysql2001mysql,mosberger1998httperf}. For the workloads of \textit{DB-S} and \textit{DB-U}, we measure the latency to select (\textit{S}) and update (\textit{U}) key-value pairs from a target database, respectively. For the other workloads, we measure the response time to fetch (\textit{F}) or upload (\textit{U}) data from/to a server. We examine both web (\textit{Apache}) and image services (\textit{ImgServer}).
We also select a write-intensive throughput-oriented workload, \textit{Ungzip}, which is used by a GNU application \cite{gailly2010gnu}.} 
In our evaluations, we \emph{co-run} latency-critical and write-intensive workloads for 30 seconds to simulate a real-world server environment.

\subsection{Overall System Performance}
\begin{figure}
	\centering
	\vspace{15pt}	
	\includegraphics[width=1\linewidth]{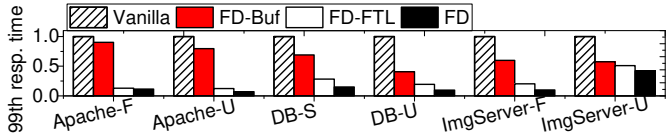}
	\vspace{-15pt}
	\caption{Response time analysis of multiple co-located workload execution. The results are normalized to \texttt{Vanilla}. \vspace{-10pt}}
	\label{fig:lat-99th}
\end{figure}

Figure \ref{fig:lat-99th} plots the $99^{th}$ percentile response time of the latency-critical applications, when co-running the latency-critical applications and write-intensive workloads under four different system configurations. Overall, \texttt{FD-Buf}, \texttt{FD-FTL}, and \texttt{FD} achieves 34\%, 76\%, and 84\% shorter $99^{th}$ percentile response latency, respectively, in comparison to \texttt{Vanilla}.

\texttt{FD-Buf} identifies page victimization via holistic NVMe storage stack optimization and cleans up SSD internal buffer in advance to accommodate I/Os of page victimization. As a result, compared to \texttt{Vanilla}, \texttt{FD-Buf} can isolate the page victimization from read I/Os. This in turn reduces the delay that user applications experience when waiting for I/O response. Nonetheless, the limited SSD internal buffer can overflow when the Linux page cache intensively evicts dirty pages. Compared to \texttt{FD-Buf}, \texttt{FD-FTL} can allocate low-latency LSB pages of NAND flash as write buffer to accommodate a large number of dirty pages. Therefore, \texttt{FD-FTL} can reduce the time during which page victimization blocks other I/O requests. 
\newedit{\texttt{FD} can further reduce the $99^{th}$ percentile response time of the evaluated latency-critical workloads by 37\%, on average, compared to \texttt{FD-FTL}. This is because \texttt{FD} adaptively suspends Linux page cache from evicting dirty pages when the SSD write buffer is full.}


\begin{figure}
\centering
\def\subfigcapskip{0pt}
\subfloat[Response time analysis.]{\label{fig:TSA5-new}\rotatebox{0}{\includegraphics[width=0.5\linewidth]{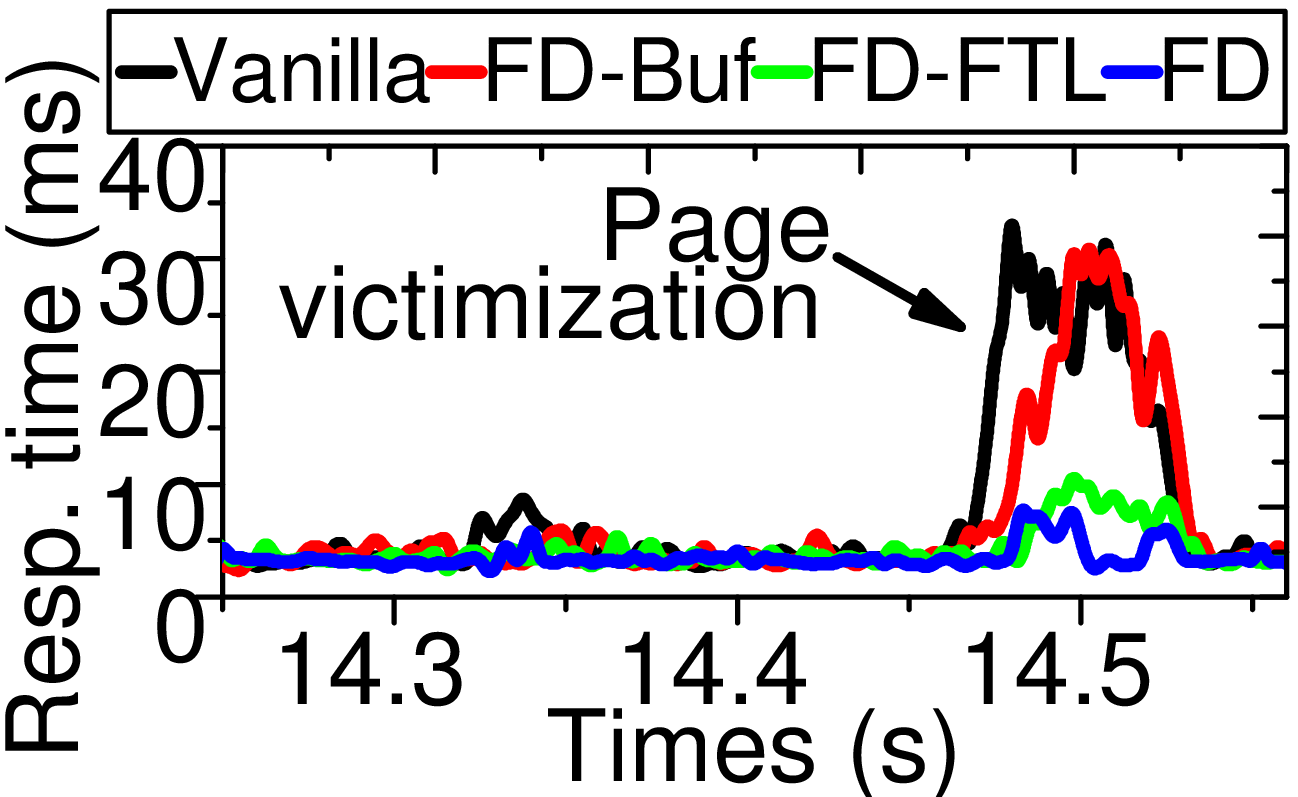}}}
\hspace{3pt}
\subfloat[Flash page wastage in \texttt{FD}.]{\label{fig:wasted_fig}\rotatebox{0}{\includegraphics[width=0.46\linewidth]{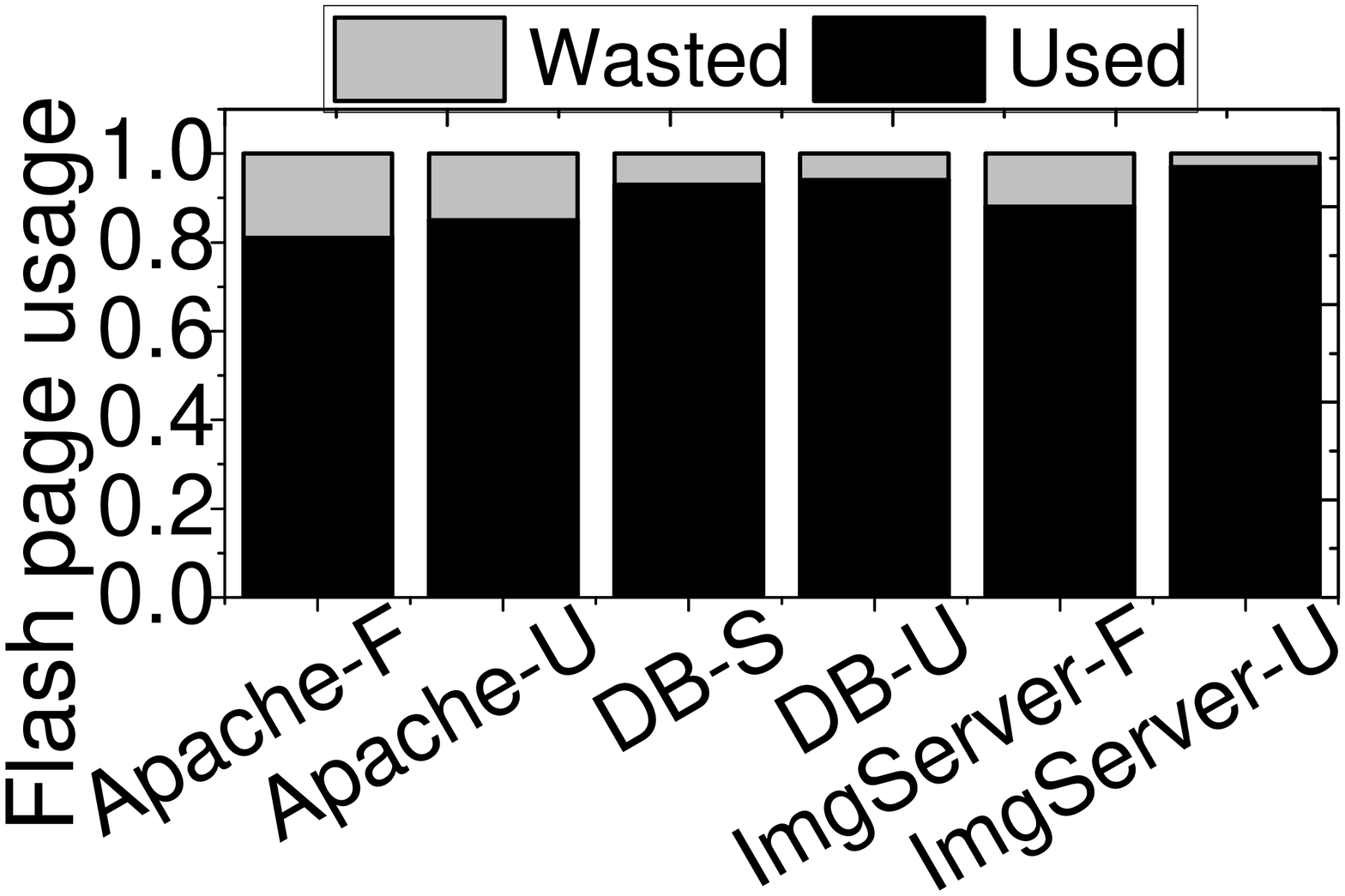}}}
\vspace{-10pt}
\caption{Response time analysis when co-running workloads \textit{Apache-U} and \textit{Ungzip}, and flash page wastage in \texttt{FD}. \vspace{-15pt}}
\label{fig:TSA12}
\end{figure}

\subsection{Performance and overhead analysis}
\noindent \textbf{Response time analysis.} Figure \ref{fig:TSA5-new} shows the response time of \textit{Apache-U} when co-running workloads \textit{Apache-U} and \textit{Ungzip}. Specifically, \texttt{Vanilla} achieves the average response time under 5 ms when there is no page victimization. Once write I/Os of page victimization arrive in the storage, the response time increases up to 35 ms. 
As \texttt{FD-Buf} can identify page victimization, it cleans up the SSD internal buffer in advance for the incoming write I/Os. While the SSD internal buffer is sufficient to accommodate all write I/Os without accessing the underlying flash media at the start of page victimization, such promising performance of \texttt{FD-Buf} unfortunately is not sustainable after serving a large number of write I/Os. 
This is because \texttt{FD-Buf} needs to evict a large number of dirty pages to the flash media, when the SSD internal buffer is full. As \texttt{FD-FTL} can mitigate the flash write latency by placing evicted data in flash LSB pages, it reduces the worst response time by 3$\times$ compared to \texttt{FD-Buf}. \texttt{FD} further reduces the long tail response time by throttling the write I/Os during page victimization events.
\newedit{Note that \texttt{FD} is also beneficial to reduce the response time when there is no page victimization event. This is because our FTL design monitors the SSD internal traffic and schedules the background data eviction of the SSD internal buffer only when the flash resources are available.}  

\noindent \textbf{Overhead analysis.} Figure \ref{fig:wasted_fig} compares the number of skipped CSB/MSB pages (\texttt{Wasted}) with the number of programmed flash pages (\texttt{Used}).
As shown in the figure, our latency-aware FTL design, on average, can utilize 90\% of flash pages in the evaluated workloads, which is much higher than the prior work (cf. Figure \ref{fig:motiv7}). This is because our proposed FTL design skips programming data in CSB/MSB pages only during the foreground eviction of the SSD internal buffer. In addition, our FTL design spreads the foreground eviction to LSB pages across flash blocks to minimize the wastage of CSB/MSB pages.

\section{Related Work}
\label{sec:related}
Several studies \cite{kim2017enlightening, yang2015split} have tried to holistically optimize the OS and storage stack to fully reap the benefits of SSD-enabled systems. These storage-stack optimizations mainly detect I/O dependencies among I/O requests, which are performed in the background and address their interferences. 
Even though the overall system performance can be improved with this optimization, they are unsuitable in scenarios that do no exhibit any I/O dependency. 
Compared to these prior works, in this work, we first reveal that OS page cache has a major influence on performance and blocks other I/O requests that have no I/O dependency with incoming requests from different user applications. 
FastDrain holistically optimizes the software and hardware modules from the OS to the underlying SSDs, which can eliminate the critical performance bottleneck of OS page cache in the existing storage stack. 

\section{Conclusion}
\label{sec:conclusion}
We observe that page victimization can degrade the user experiences. 
To address this, we propose a co-design of OS kernel and flash firmware, which can mitigate the overhead of page victimization.
\newedit{Our evaluation results demonstrate that the proposed approach outperforms the conventional system by 84\%, in terms of the $99^{th}$ percentile response time.}

\section{Acknowledgements}
We thank anonymous reviewers for their constructive feedback. This research is mainly supported by NRF 2016R1C1B2015312, DOE DE-AC02-05CH 11231, KAIST Start-Up Grant (G01190015), and MemRay grant (G01190170). N.S. Kim is supported in part by grants from NSF CNS-1705047. M. Kandemir is supported in part by NSF grants 1822923, 1439021, 1629915, 1626251, 1629129, 1763681, 1526750 and 1439057. Myoungsoo Jung is the corresponding author.

\ifCLASSOPTIONcaptionsoff
  \newpage
\fi

{
\bibliographystyle{abbrv}
\bibliography{nvme-gem5}
}

\end{document}